\documentclass{aa}  
\usepackage{graphicx}
\usepackage{txfonts}
\usepackage{xcolor}
\usepackage{hyperref}
\hypersetup{colorlinks, linkcolor=blue, citecolor=blue, urlcolor=blue}

\newcommand{\te}{t_{\rm E}}
\newcommand{\thetae}{\theta_{\rm E}}

\newcommand{\pirel}{\pi_{\rm rel}}
\newcommand{\pie}{\pi_{\rm E}}

\newcommand{\dl}{D_{\rm L}}
\newcommand{\ds}{D_{\rm S}}

\definecolor{brown}{rgb}{0.59, 0.29, 0.0}
\definecolor{darkgreen}{rgb}{0.0, 0.42, 0.24}
\definecolor{darkblue}{rgb}{0.01, 0.31, 0.59}
\definecolor{darkblue}{rgb}{0.0, 0.25, 0.42}
\definecolor{blue}{rgb}{0.0,0.0,1.0}
\definecolor{green}{rgb}{0.0,1.0,0.0}

\def\eqalign#1{\null\,\vcenter{\openup\jot
        \ialign{\strut\hfil$\displaystyle{##}$&$
        \displaystyle{{}##}$\hfil \crcr#1\crcr}}\,}

\begin{document}

\title{KMT-2024-BLG-0404L: A triple microlensing system consisting of a star, a brown dwarf, and a planet}
\titlerunning{KMT-2024-BLG-0404L}

\author{
     Cheongho~Han\inst{\ref{inst1}} 
\and Andrzej~Udalski\inst{\ref{inst2}} 
\and Chung-Uk~Lee\inst{\ref{inst3}\thanks{\tt leecu@kasi.re.kr}}
\\
(Leading authors)
\\
     Yoon-Hyun~Ryu\inst{\ref{inst3}} 
\and Michael~D.~Albrow\inst{\ref{inst4}}   
\and Sun-Ju~Chung\inst{\ref{inst3}}      
\and Andrew~Gould\inst{\ref{inst5},\ref{inst6}}      
\and Kyu-Ha~Hwang\inst{\ref{inst3}} 
\and Youn~Kil~Jung\inst{\ref{inst3}} 
\and In-Gu~Shin\inst{\ref{inst7}} 
\and Yossi~Shvartzvald\inst{\ref{inst8}}   
\and Jennifer~C.~Yee\inst{\ref{inst7}}   
\and Hongjing~Yang\inst{\ref{inst9}}     
\and Weicheng~Zang\inst{\ref{inst7},\ref{inst9}}     
\and Doeon~Kim\inst{\ref{inst1}}
\and Dong-Jin~Kim\inst{\ref{inst3}} 
\and Byeong-Gon~Park\inst{\ref{inst3}} 
\and Richard~W.~Pogge\inst{\ref{inst6},\ref{inst10}}
\\
(The KMTNet Collaboration)
\\
     Przemek~Mr{\'o}z\inst{\ref{inst2}} 
\and Micha{\l}~K.~Szyma{\'n}ski\inst{\ref{inst2}}
\and Jan~Skowron\inst{\ref{inst2}}
\and Rados{\l}aw~Poleski\inst{\ref{inst2}} 
\and Igor~Soszy{\'n}ski\inst{\ref{inst2}}
\and Pawe{\l}~Pietrukowicz\inst{\ref{inst2}}
\and Szymon~Koz{\l}owski\inst{\ref{inst2}} 
\and Krzysztof~A.~Rybicki\inst{\ref{inst2},\ref{inst8}}
\and Patryk~Iwanek\inst{\ref{inst2}}
\and Krzysztof~Ulaczyk\inst{\ref{inst11}}
\and Marcin~Wrona\inst{\ref{inst2},\ref{inst12}}
\and Mariusz~Gromadzki\inst{\ref{inst2}}          
\and Mateusz~J.~Mr{\'o}z\inst{\ref{inst2}} 
\and Micha{\l} Jaroszy{\'n}ski\inst{\ref{inst2}}
\\
(The OGLE Collaboration)
}

\institute{
      Department of Physics, Chungbuk National University, Cheongju 28644, Republic of Korea                                  \label{inst1}    
\and  Astronomical Observatory, University of Warsaw, Al.~Ujazdowskie 4, 00-478 Warszawa, Poland                              \label{inst2}   
\and  Korea Astronomy and Space Science Institute, Daejon 34055, Republic of Korea                                            \label{inst3}  
\and  University of Canterbury, Department of Physics and Astronomy, Private Bag 4800, Christchurch 8020, New Zealand         \label{inst4}  
\and  Max Planck Institute for Astronomy, K\"onigstuhl 17, D-69117 Heidelberg, Germany                                        \label{inst5}  
\and  Department of Astronomy, The Ohio State University, 140 W. 18th Ave., Columbus, OH 43210, USA                           \label{inst6}  
\and  Center for Astrophysics $|$ Harvard \& Smithsonian 60 Garden St., Cambridge, MA 02138, USA                              \label{inst7}  
\and  Department of Particle Physics and Astrophysics, Weizmann Institute of Science, Rehovot 76100, Israel                   \label{inst8}  
\and  Department of Astronomy, Tsinghua University, Beijing 100084, China                                                     \label{inst9}  
\and  Center for Cosmology and AstroParticle Physics, Ohio State University, 191 West Woodruff Ave., Columbus, OH 43210, USA  \label{inst10} 
\and  Department of Physics, University of Warwick, Gibbet Hill Road, Coventry, CV4 7AL, UK                                   \label{inst11}
\and  Villanova University, Department of Astrophysics and Planetary Sciences, 800 Lancaster Ave., Villanova, PA 19085, USA   \label{inst12} 
}                                                                                                                                                       
\date{Received ; accepted}

\abstract
{}
{
We have investigated the lensing event KMT-2024-BLG-0404. The light curve of the event 
exhibited a complex structure with multiple distinct features, including two prominent 
caustic spikes, two cusp bumps, and a brief discontinuous feature between the caustic 
spikes. While a binary-lens model captured the general anomaly pattern, it could not 
account for a discontinuous anomaly feature between the two caustic spikes. 	
}
{
To explore the origin of the unexplained feature, we conducted more advanced modeling 
beyond the standard binary-lens framework.  This investigation demonstrated that the 
previously unexplained anomaly was resolved by introducing an additional lens component 
with planetary mass.
}
{
The estimated masses of the lens components are 
$M_{\rm p}= 17.3^{+25.5}_{-8.8}~M_{\rm E}$ for the planet, and 
$M_{\rm h,A}=0.090^{+0.133}_{-0.046}~M_\odot$ and 
$M_{\rm h,B}=0.026^{+0.038}_{-0.013}~M_\odot$ for the binary host stars. 
Based on these mass estimates, the lens system is identified as a planetary system where 
a Uranus-mass planet orbits a binary consisting of a late M dwarf and a brown dwarf.  The 
distance to the planetary system is estimated to be $\dl = 7.21^{+0.93}_{-0.97}$~kpc, with 
an 82\% probability that it resides in the Galactic bulge.  This discovery represents the 
ninth planetary system found through microlensing with a planet orbiting a binary host. 
Notably, it is the first case where the host consists of both a star and a brown dwarf.  
} 
{}


\maketitle

\section{Introduction}\label{sec:one}

Microlensing is a technique that can detect planets orbiting around binary stars. 
Microlensing discoveries of these planets are possible because both the planetary 
companion and the binary companion to the primary lens produce distinct signals. 
Consequently, the presence of a planet and the binary nature of the host can be 
identified for a fraction of events in which the source passes through the perturbation 
region created by both lens companions. \citet{Han2008} was the first to propose the 
potential of using the microlensing technique to detect planets in binary star systems. 
Since that time,  eight microlensing planetary systems within binaries have been 
reported.

The first microlensing planet in a binary system was discovered by \citet{Gould2014} 
through the analysis of the OGLE-2013-BLG-0341 lensing event. The binary nature of 
the lens was recognized from an anomaly near the peak of the event's light curve and 
a small bump that appeared well before the peak. The planetary signal was detected 
as a dip in the light curve on the rising side of the event. The planet, with a mass 
approximately twice that of Earth, is located at a projected separation comparable 
to the distance between the Earth and the Sun. Its host star orbits a slightly more 
massive companion with a projected separation of about 10 to 15 AU.

The second microlensing planet in a binary system was discovered through the analysis 
of the lensing event OGLE-2008-BLG-092 by \citet{Poleski2014}. In this microlensing 
event, distinct signals from both the binary stellar and planetary companions to the 
primary were identified, each explainable by the light curve of a single-mass object. 
The planet, which has a mass approximately four times that of Uranus, orbits a binary 
system composed of a K dwarf and an M dwarf.

The third microlensing planet in a binary was identified by \citet{Bennett2016} through 
their analysis of the lensing event OGLE-2007-BLG-349.  The peak part of the lensing 
light curve in this event clearly indicated a signal from a planet, while an additional 
signal hinted at the presence of another lensing mass.  Their analysis provided two 
plausible model interpretations: one suggesting two planets orbiting a single star, 
and the other proposing a circumbinary planet. The ambiguity between these models was 
resolved by detecting excess flux from the binary lens star system using {\it Hubble 
Space Telescope} imaging. The planet identified is a gas giant orbiting a pair of M 
dwarfs.

\citet{Han2017} reported the discovery of the fourth planet in a binary system 
through the analysis of the lensing event OGLE-2016-BLG-0613. The light curve 
displayed the typical features of a caustic-crossing binary-lens event, with two 
caustic spikes and a U-shaped trough between them.  However, an additional 
short-term, discontinuous feature appeared within the trough. A detailed analysis 
uncovered three possible models, all suggesting that this extra feature was produced 
by a planetary companion to the binary lens.  The primary star's mass was most likely 
around $0.7~M_\odot$, and the planet was classified as a super-Jupiter, although the 
mass of the binary companion varies depending on the models.

The fifth planetary system, OGLE-2006-BLG-284L, was reported by \citet{Bennett2020} 
following a detailed analysis of the microlensing event OGLE-2006-BLG-284. In this 
event, the planetary signal appeared as a brief but distinct deviation from the 
typical two-spike light curve seen in a caustic-crossing binary-lens event. A thorough 
analysis confirmed that this short-term signal was caused by a planetary companion. 
The lens system consists of a giant planet orbiting a pair of M dwarf stars. This 
discovery further highlighted microlensing's ability to detect planetary companions 
in binary star systems.

The sixth microlensing planet in a binary system was discovered through the analysis 
of the lensing light curve of OGLE-2018-BLG-1700. In this event, the signals from 
both the stellar and planetary companions overlapped in the central region, leading 
to a complex anomaly in the peak region of the light curve. \citet{Han2020} 
characterized the triple nature of the lens by separating the anomaly pattern into two 
distinct components, each corresponding to a binary-lens event.  One binary lens pair 
exhibited a mass ratio of approximately 0.01 between the lens objects, suggesting the 
companion is a planet, while the other had a mass ratio of about 0.3.  Through Bayesian 
analysis, they estimated the mass of the planetary companion to be around four times 
that of Jupiter, with the stellar binary components having masses of approximately 
$0.4~M_\odot$ and $0.12~M_\odot$.

The seventh triple-lens system consisting of a planet and a binary was identified by 
\citet{Zang2021} through the analysis of central anomalies in the lensing event 
KMT-2020-BLG-0414.  They found that the planet has an extremely low mass, similar to 
that of Earth.  From a subsequent followup observation using Keck Adaptive Optics, 
\citet{Zhang2024} found that the planet is hosted by a binary composed of a brown 
dwarf and a white dwarf, as evidenced by the non-detection of the lens flux.

The last microlensing planet in a binary stellar system was discovered by \citet{Han2024} 
through their analysis of the lensing event OGLE-2023-BLG-0836. Similar to the 
event OGLE-2018-BLG-1700, the anomaly displayed central deviations of complex anomaly 
pattern that could not be explained by a binary-lens model. This anomaly could be 
explained by a triple lens model in which one member of the lens is a planet.  The 
planet has a mass of 4.4 times that of Jupiter, and the stellar binary consists of 
two stars with masses of approximately $0.7~M_\odot$ and $0.6~M_\odot$.

Interpreting the lensing behavior of a planet in a binary system requires modeling 
a triple-lens system. The lensing behave of triple-lens system was first introduced 
by \citet{Grieger1989}, who explored the impact of shear caused by a third body 
and demonstrated the resulting complexity in the caustic structure. \citet{Rhie2002} 
later showed that the lens equation for a triple-lens system is a two-dimensional 
vector equation, which can be embedded into a tenth-order analytic polynomial equation 
in one complex variable. \citet{Han2001} demonstrated that the anomalies induced by 
multiple planets could be approximated by superimposing the anomalies from individual 
planets. The first triple-lens system, OGLE-2006-BLG-109L, consisting of two planets 
orbiting a low-mass star \citep{Gaudi2008}, was interpreted through this approach. 
The complexities of modeling a triple-lens system were thoroughly examined by 
\citet{Danek2015} and \citet{Danek2019}.  These complexities were illustrated from 
the analysis of the lensing event OGLE-2013-BLG-0723, where the lens responsible for 
the light curve was initially interpreted as a triple system, including a Venus, a 
brown dwarf, and a star \citep{Udalski2015b}, but later reinterpreted as a rotating 
binary system \citep{Han2016}.

We conducted a project involving a detailed analysis of anomalous microlensing events 
identified by the Korea Microlensing Telescope Network (KMTNet) survey \citep{Kim2016}. 
The aim of the project is to investigate events with anomalies in their light curves 
that cannot be explained by the conventional binary-lens single-source (2L1S) or 
single-lens binary-source (1L2S) models. Through this analysis, we identified cases 
requiring more advanced modeling beyond the standard 2L1S or 1L2S frameworks. In most 
cases, these complex anomalies were resolved by introducing additional lens or source 
components.  Previous events with light curves interpreted using four-body (lens+source) 
models are listed in Table 1 of \citet{Han2023}.  In this paper, we report an additional 
planet in a binary system, discovered through a systematic investigation of KMTNet data 
from the 2024 season.

\section{Observation and data} \label{sec:two}

The planetary system was discovered through a detailed analysis of the microlensing event 
KMT-2024-BLG-0404. The KMTNet group detected this event on April 3, 2024, corresponding 
to the abbreviated Heliocentric Julian date ${\rm HJD}^\prime \equiv {\rm HJD} - 2460000 = 
403$. The event's baseline $I$-band magnitude was measured at $I_{\rm base} = 18.63$. The 
source is located at equatorial coordinates $({\rm RA}, {\rm DEC})_{\rm J2000} =$ (18:04:35.30, 
-27:57:12.60), corresponding to Galactic coordinates $(l, b) = (2^\circ\hskip-2pt.9528, 
-3^\circ\hskip-2pt.1045)$. The source lies in the overlapping region of KMTNet prime fields
BLG03 and BLG43. However, data from the BLG03 field were unavailable, as the source fell 
into the gap between the four camera chips.  The event was also independently detected and 
announced on April 11, 2024 by the Optical Gravitational Lensing Experiment group 
\citep[OGLE;][]{Udalski2015}, which designated it as OGLE-2024-BLG-0378.

In our analysis, we utilized combined data from both the KMTNet and OGLE surveys. KMTNet 
data were collected using three identical 1.6-meter telescopes located on three different 
continents in the Southern Hemisphere, allowing for continuous monitoring of microlensing 
events. These telescopes are situated at Cerro Tololo Inter-American Observatory in Chile 
(KMTC), Siding Spring Observatory in Australia (KMTA), and South African Astronomical 
Observatory (KMTS). Each telescope is equipped with a camera covering a 4-square-degree 
field of view.  Data from the OGLE survey were acquired using the 1.3-meter Warsaw telescope 
at Las Campanas Observatory in Chile, which features a 1.4-square-degree field of view. 
Both KMTNet and OGLE primarily observed in the $I$ band, with approximately 10\% of images 
taken in the $V$ band for source color measurement.

The reduction of images and photometry for the lensing event was carried out using 
automated pipelines specific to each survey. For the KMTNet survey, image reduction was 
performed using the pipeline developed by \citet{Albrow2000}, while the OGLE survey used 
the pipeline developed by \citet{Udalski2003}. To ensure optimal data quality, we conducted 
additional photometric analysis on the KMTNet data using a code developed by \citet{Yang2024}. 
For both data sets, we rescaled the error bars to match the data scatter and normalized 
the $\chi^2$ per degree of freedom to unity, following the procedure described by 
\citet{Yee2012}.

\begin{figure}[t]
\includegraphics[width=\columnwidth]{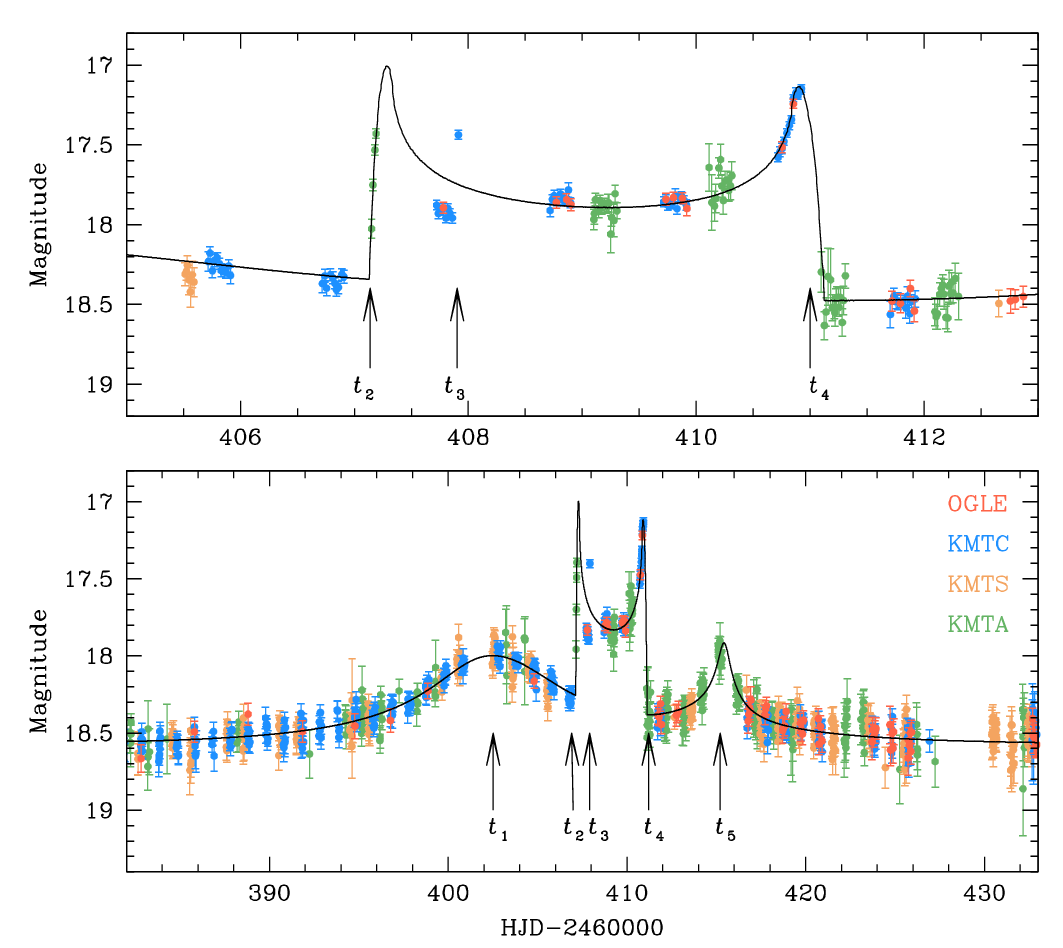}
\caption{
Light curve of the lensing event KMT-2024-BLG-0404. The lower panel presents the overall view, 
while the upper panel provides a close-up of the caustic-crossing features. The curve overlaid 
on the data points represents the best-fit 2L1S model, which was derived by excluding data from 
the interval $407.5 < {\rm HJD}^\prime < 408.0$.  In the lower panel, arrows labeled $t_1$, 
$t_2$, $t_3$, $t_4$, and $t_5$ mark the positions of specific anomaly features.
}
\label{fig:one}
\end{figure}

\section{Light curve analysis} \label{sec:three}

Figure~\ref{fig:one} presents the lensing light curve for the event KMT-2024-BLG-0404, combining 
data from the OGLE and KMTNet surveys. The light curve displays a complex pattern with multiple 
anomaly features. First, the sharp rise and fall at times $t_2$ and $t_4$ correspond to a pair 
of spikes caused by the source crossing a caustic. Second, the smooth rise and fall at times 
$t_1$ and $t_5$ appear as bumps, likely due to the source approaching a caustic cusp. Additionally, 
a notable feature near $t_3$ likely involves a caustic, as indicated by the abrupt change in 
magnitude in the data between ${\rm HJD}^\prime = 407.7$ and 407.9.  In a two-mass lens system, 
the caustic forms a closed curve, and in the light curve of a caustic-crossing 2L1S event, the 
region between the two spikes typically forms a smooth "U"-shaped profile. The discontinuity at 
$t_3$ in the trough strongly suggests the presence of a third mass in the lens system.

\begin{table}[t]
\caption{Lensing parameters of the 2L1S model.\label{table:one}}
\begin{tabular*}{\columnwidth}{@{\extracolsep{\fill}}lllll}
\hline\hline
\multicolumn{1}{c}{Parameter}         &
\multicolumn{1}{c}{Value}             \\
\hline
 $t_0$ (HJD$^\prime$)       &  $412.874 \pm 0.063$     \\
 $u_0$                      &  $0.2449 \pm 0.0015$     \\
 $\te$ (days)               &  $11.265 \pm 0.093 $     \\
 $s$                        &  $1.5537 \pm 0.0047$     \\
 $q$                        &  $3.45 \pm 0.10    $     \\
 $\alpha$ (rad)             &  $5.6777 \pm 0.0082$     \\
 $\rho$ (10$^{-3}$)         &  $5.02 \pm 0.12    $     \\
\hline             
\end{tabular*}
\tablefoot{ ${\rm HJD}^\prime = {\rm HJD}- 2460000$.  }
\end{table}

\subsection{Binary lens model} \label{sec:three-one}

To analyze the caustic-related features, we began with a 2L1S model analysis of the light
curve, aiming to determine the lensing solution -- a set of parameters that best explains 
the observed light curve. A 2L1S event is characterized by seven fundamental parameters. 
Three of these parameters describe the source's approach to the lens: $t_0$ represents the 
time of closest approach of the source to the lens, $u_0$ is the projected separation between 
the source and the lens at that moment, scaled to the angular Einstein radius ($\thetae$), 
and $\te$ is the event timescale, defined as the time it takes for the source to cross 
$\thetae$.  Two additional parameters describe the binary lens: $s$, the projected separation 
between the lens components ($M_1$ and $M_2$), scaled by $\thetae$, and $q$, the mass ratio 
between the two components. The parameter $\alpha$ indicates the angle of the source's 
trajectory relative to the binary axis. Finally, $\rho$, defined as the ratio of the source 
radius ($\theta_*$) to the angular Einstein radius, accounts for finite-source effects, 
which lead to magnification attenuation during caustic crossings or approaches.

The modeling process employed a combination of grid and downhill methods.  Due to the 
complexity of the $\chi^2$ surface for the 2L1S lensing parameters, relying solely on a 
downhill approach can be challenging.  To tackle this, we conducted modeling with $(s, q)$ 
as grid parameters, employing multiple starting values for $\alpha$, while optimizing the 
remaining parameters using a downhill approach.  Setting $(s, q)$ as grid parameters is 
also crucial for exploring degenerate solutions, in which similar light curves may result 
from solutions with vastly different lensing parameters. In the downhill approach, we 
employed a Markov Chain Monte Carlo (MCMC) algorithm. For the local solutions identified 
through this procedure, we subsequently refined the lensing parameters by allowing them to 
vary.

\begin{figure}[t]
\includegraphics[width=\columnwidth]{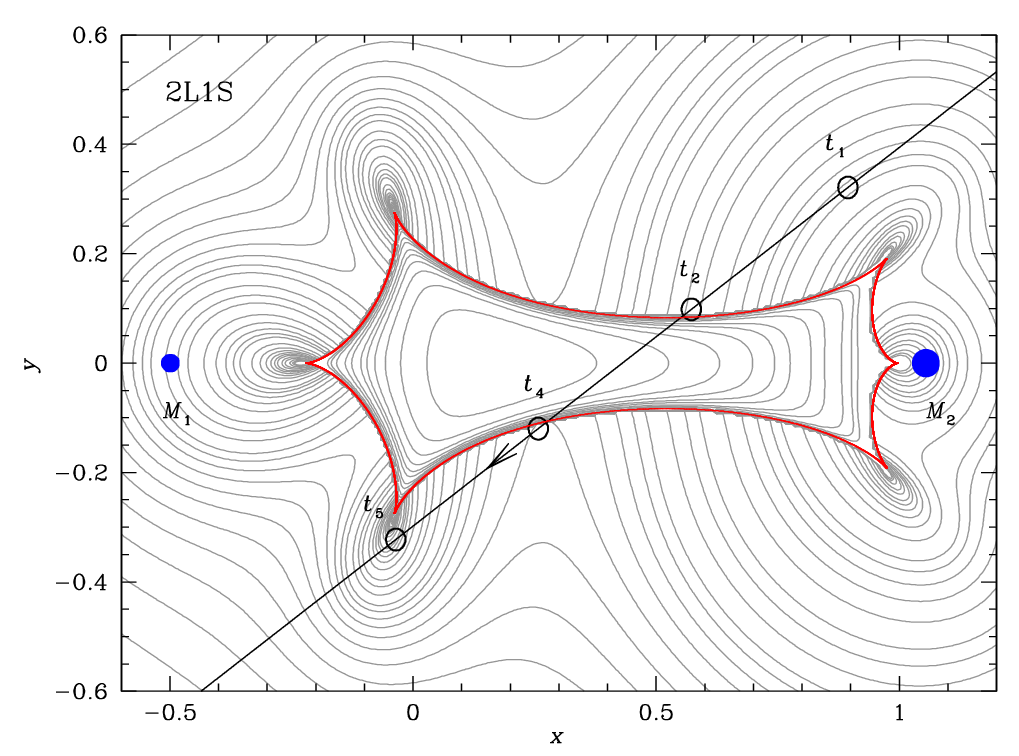}
\caption{
Lens system configuration corresponding to the best-fit 2L1S model.  The red cuspy closed figure 
represents the caustic, while the two blue dots labeled $M_1$ and $M_2$ denote the positions 
of the binary-lens components. The arrowed line illustrates the trajectory of the source.  The 
four empty circles, scaled to the source size, mark the source positions along its trajectory 
at the times corresponding to the anomaly features at $t_1$, $t_2$, $t_4$, and $t_5$. The gray 
curves surrounding the caustic represent the equi-magnification contours.
}
\label{fig:two}
\end{figure}

The 2L1S modeling yields a solution that captures the overall pattern of the anomaly, with 
the exception of the feature around $t_3$. Figure~\ref{fig:one} displays the model curve 
corresponding to the 2L1S solution derived from this analysis.  The binary parameters for 
this solution are approximately $(s, q) \sim (1.6, 3.5)$, indicating that the lens is a binary 
consisting two masses with a projected separation slightly exceeding the Einstein radius.  The 
event time scale is $\te \sim 11$ days. The time scales corresponding to the individual lens 
components are given by $t_{{\rm E},1}=[1/(1+q)]^{1/2}\te \sim 5.3$~days and $t_{{\rm E},2}=
[q/(1+q)]^{1/2}\te \sim 9.9$~days. These relatively short time scales suggest that the masses 
of the binary lens components are likely to be small. The complete set of lensing parameters 
for the model is provided in Table~\ref{table:one}.

Figure~\ref{fig:two} illustrates the configuration of the 2L1S lens system, depicting the 
source trajectory with respect to the lens components and the resulting caustic. The lens 
generates a single resonant caustic with six cusps elongated along the binary axis. The source 
passes diagonally through the caustic, creating a spike at $t_2$ upon entry and another spike 
at $t_4$ upon exit. The bumps at $t_1$ and $t_5$ occur as the source approaches the upper right 
and lower left cusps of the caustic, respectively. In the figure, we mark the positions of the 
source at the moments of these major anomaly features. However, between $t_2$ and $t_4$, there 
is no caustic feature that could account for the anomaly seen in the lensing light curve around 
$t_3$, indicating that the 2L1S model is insufficient to fully explain all the observed anomaly 
features.

\begin{figure}[t]
\includegraphics[width=\columnwidth]{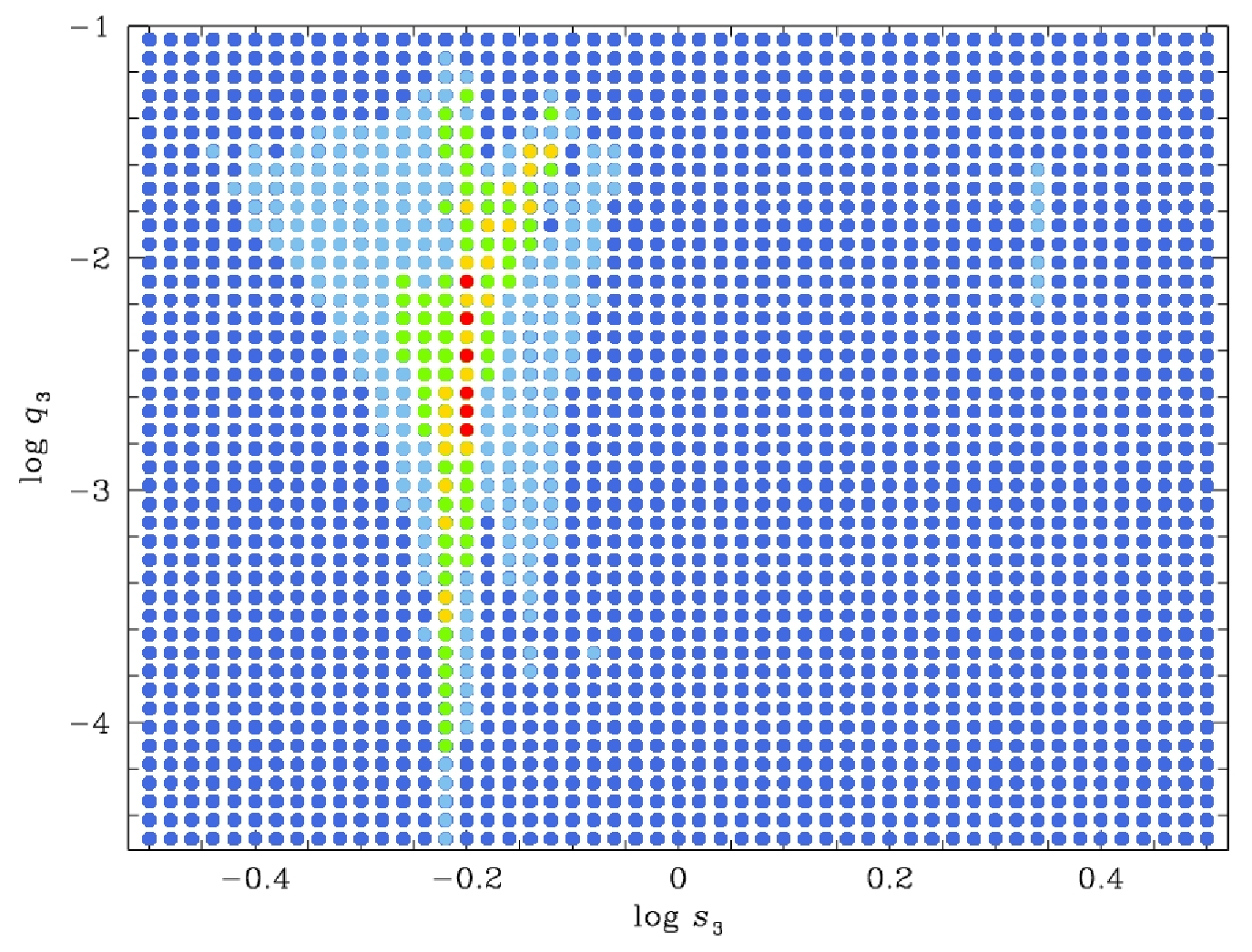}
\caption{
map of $\Delta\chi^2$ on the $\log s_3$--$\log q_3$ parameter surface derived from the 
triple-lens modeling. The color coding indicates points with $\Delta\chi^2 \leq 1n\sigma$ 
(red), $\leq 2n\sigma$ (yellow), $\leq 3n\sigma$ (green), $\leq 4n\sigma$ (cyan), and 
$\leq 5n\sigma$ (blue), where $n=4$.
}
\label{fig:three}
\end{figure}

\subsection{Triple lens model} \label{sec:three-two}

The fact that the discontinuous feature near $t_3$ cannot be explained by a 2L1S model 
suggests the presence of an additional cause for the observed anomaly. We explore two 
possible explanations. The first is the existence of an additional companion to the source, 
resulting in a system composed of two lens masses and two source stars (a 2L2S event). Such 
configurations have been observed in events like MOA-2010-BLG-117 \citep{Bennett2018}, 
KMT-2018-BLG-1743 \citep{Han2021a}, OGLE-2016-BLG-1003 \citep{Jung2017}, KMT-2019-BLG-0797 
\citep{Han2021b}, KMT-2021-BLG-1898 \citep{Han2022}, OGLE-2018-BLG-0584, and KMT-2018-BLG-2119 
\citep{Han2023}.  The second possibility involves an additional companion to the lens, leading 
to a system with three lens masses and a single source star (a 3L1S event), as illustrated by 
the events discussed in Sect.~\ref{sec:one}. Given that the anomaly around $t_3$ constitutes 
a minor perturbation, in the 3L1S scenario, the lens companion is likely to be a very low-mass 
object, while in the 2L2S scenario, the source companion would probably be a faint star.

\begin{figure}[t]
\includegraphics[width=\columnwidth]{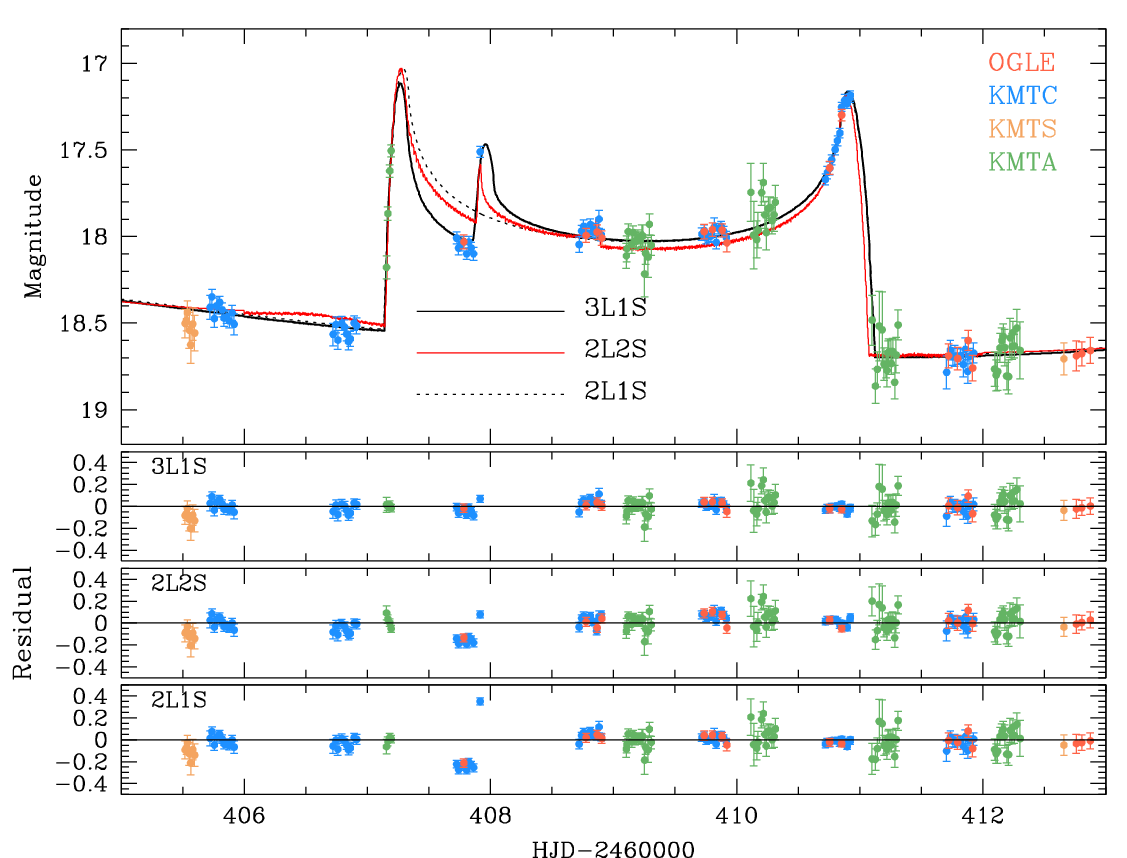}
\caption{
Comparison of the models (2L1S, 2L2S, and 3L1S) in the region around caustic spikes.
The lower panels show the residuals from the models.
}
\label{fig:four}
\end{figure}

Among the two four-body (lens + source) models tested, we reject the 2L2S interpretation.  
The primary reason is that the 2L1S model leaves both positive and negative residuals in 
the data from the night of ${\rm HJD}^\prime = 407$, whereas an additional source can only 
account for positive deviations.  As a result, while the data point at ${\rm HJD}^\prime = 
407.92$ with a positive deviation could be explained by a 2L2S model, the remaining nine 
points with negative deviations (one from OGLE and eight from KMTC) cannot.  This is 
illustrated in Figure~\ref{fig:four}, which presents the best-fit 2L2S model and its residuals.  
Given that the time interval between the two caustic spikes at $t_2$ and $t_4$ is relatively 
short (approximately 4 days) and that all features except for the one around $t_4$ are well 
described by a static model, the anomaly near $t_3$ is not attributed to lens or source 
orbital motion.

The 3L1S modeling was conducted in two phases. In the first phase, we performed a grid search 
focused on the parameters related to the third lens mass ($M_3$), while keeping the parameters 
for the other two lens masses fixed to the values determined from the 2L1S modeling. The three 
parameters for $M_3$ are $(s_3, q_3)$, which represent the projected separation and mass ratio 
between $M_3$ and $M_1$, and $\psi$, which indicates the orientation of $M_3$ relative to the 
$M_1$--$M_2$ axis. This method is viable because the 2L1S model well describes the overall light 
curve pattern, allowing the anomaly near $t_3$ to be treated as a perturbation. In the second 
phase, we refined the lensing parameters for the local solutions identified from the $\chi^2$ 
surface of the grid search.

Figure~\ref{fig:three} shows the $\Delta\chi^2$ map on the $\log s_3$--$\log q_3$ plane, 
revealing a distinct local minimum. The complete set of lensing parameters for the refined 
3L1S model is listed in Table~\ref{table:two}. The parameters for the second lens component 
remain nearly identical to those of the 2L1S model, indicating that the third mass causes 
only a small perturbation to the lensing system. The derived parameters for the third mass 
are approximately $(s_3, q_3) \sim (0.63, 1.9 \times 10^{-3})$, suggesting that the third 
body is a planetary-mass object located within the Einstein ring of the host binary.  The 
planet is positioned between the two binary components, with an orientation angle of 
-17.48 degrees measured at the location of $M_1$ relative to the $M_1$--$M_2$ axis.

\begin{figure}[t]
\includegraphics[width=\columnwidth]{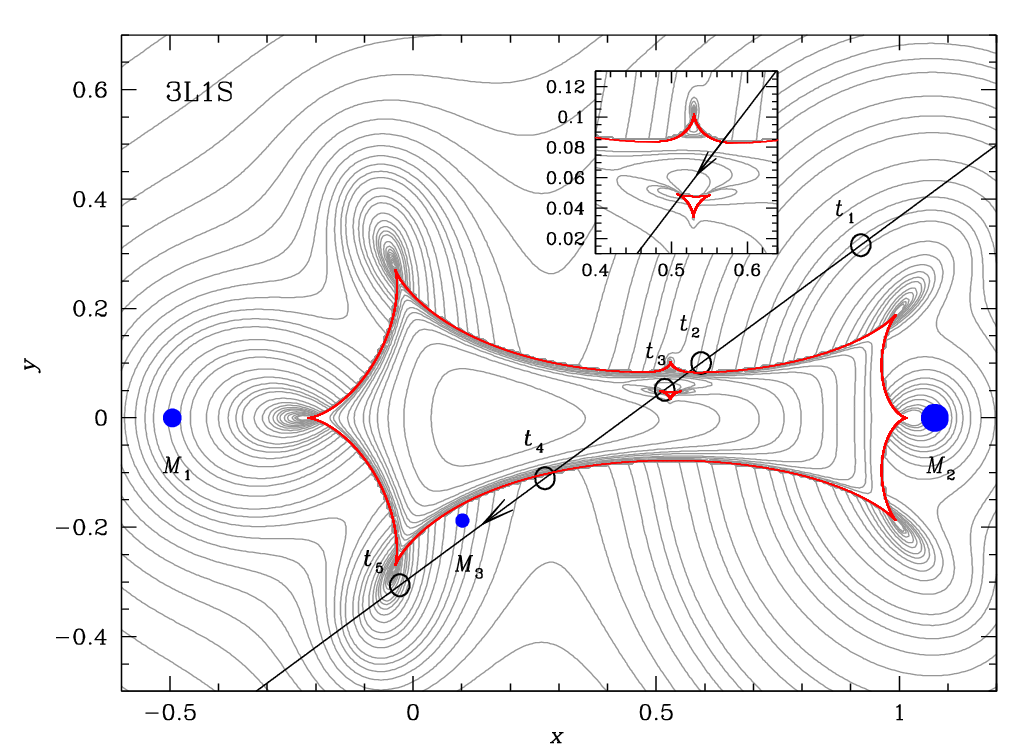}
\caption{
Lens system configuration for the 3L1S model.  Notations are same as those in Fig.~\ref{fig:two}. 
The inset presents a close-up of the area surrounding the planet-induced caustics. In comparison 
to the configuration shown in Fig.~\ref{fig:two}, the position of the third lens component ($M_3$) 
and the source position at time $t_3$ are additionally indicated.
}
\label{fig:five}
\end{figure}

Figure~\ref{fig:four} shows the model curve for the 3L1S solution in the region around the 
two caustic spikes.  For comparison, the 2L1S model curve is also displayed (dotted line). 
It shows that the 3L1S model well describes the anomaly feature around $t_3$, which could 
not be explained by the 2L1S model. The lens system configuration for the 3L1S solution is 
illustrated in Figure~\ref{fig:five}. The configuration appears similar to that of the 2L1S 
solution, as expected from the similarity in lensing parameters, except for those related to 
$M_3$. The third lens component generates a small caustic that distorts the upper fold of the 
main caustic induced by the binary.  The planet-induced caustic resembles a pair of peripheral 
caustics created by a planet with a separation less than the Einstein radius \citep{Han2006}.  
In this case, the region between the pair of peripheral caustics exhibits negative deviations.  
The observed negative deviations around $t_3$ occurred as the source passed through this area.

To illustrate the correlations among the lensing parameters, we present scatter plots of 
the MCMC chain points for the 2L1S and 3L1S models in Figures~\ref{fig:a1} and \ref{fig:a2}, 
respectively.

\begin{table}[t]
\caption{Lensing parameters of the 3L1S model.\label{table:two}}
\begin{tabular*}{\columnwidth}{@{\extracolsep{\fill}}lllll}
\hline\hline
\multicolumn{1}{c}{Parameter}         &
\multicolumn{1}{c}{Value}             \\
\hline
 $t_0$ (HJD$^\prime$)       &   $413.074 \pm 0.069 $  \\
 $u_0$                      &   $0.2409 \pm 0.0017 $  \\
 $\te$ (days)               &   $11.243 \pm 0.093  $  \\
 $s_2$                      &   $1.5681 \pm 0.0051 $  \\
 $q_2$                      &   $3.46 \pm 0.12     $  \\
 $\alpha$ (rad)             &   $5.7028 \pm 0.0095 $  \\
 $s_3$                      &   $0.6253 \pm 0.0081 $  \\
 $q_3$ (10$^{-3}$)          &   $1.93 \pm 0.14     $  \\
 $\psi$ (rad)               &   $-0.3051 \pm 0.0067$  \\
 $\rho$ (10$^{-3}$)         &   $4.69 \pm 0.14     $  \\
\hline             
\end{tabular*}
\end{table}

\section{Source star and angular Einstein radius} \label{sec:four}

We specify the source star of the event by measuring its dereddened color and magnitude.
Specifying the source star is crucial not only for fully characterizing the event, including 
the source, but also for measuring the angular Einstein radius. The Einstein radius is 
estimated as
\begin{equation}
\thetae = { \theta_* \over \rho},
\label{eq1}
\end{equation}
where the angular source radius ($\theta_*$) is inferred from the color and magnitude, and the
normalized source radius is obtained from the modeling. For KMT-2024-BLG-0404, the value of
$\rho$ was measured from the analysis of the data during caustic crossings at $t_2$, $t_3$, 
and $t_4$.

In order to estimate the reddening-corrected source color and magnitude, $(V-I, I)_0$, from its
instrumental values, $(V-I, I)$, we used the \citet{Yoo2004} method, which utilizes the centroid
of red giant clump (RGC) in the color-magnitude (CMD) as a reference for calibration 
\citep{Rattenbury2007}. According to the method, we first measured $(V-I, I)$ by regressing the 
$V$ and $I$-band photometry data processed using the pyDIA photometry code \citep{Albrow2017} 
with respect to the model light curve. We then positioned the source in the instrumental CMD, 
constructed using the same pyDIA code, for stars located near the source in the KMTC image. The 
dereddened values of the source color and magnitude were then estimated as
\begin{equation}
(V-I, I)_0 = (V-I, I)_{{\rm RGC},0} + \Delta(V-I, I),
\label{eq2}
\end{equation}
where $(V-I, I)_{{\rm RGC},0}$ denote the dereddened color and magnitude of the RGC centroid
and $\Delta(V-I, I)$ denote the offsets in color and magnitude of the source from the RGC centroid.

\begin{figure}[t]
\includegraphics[width=\columnwidth]{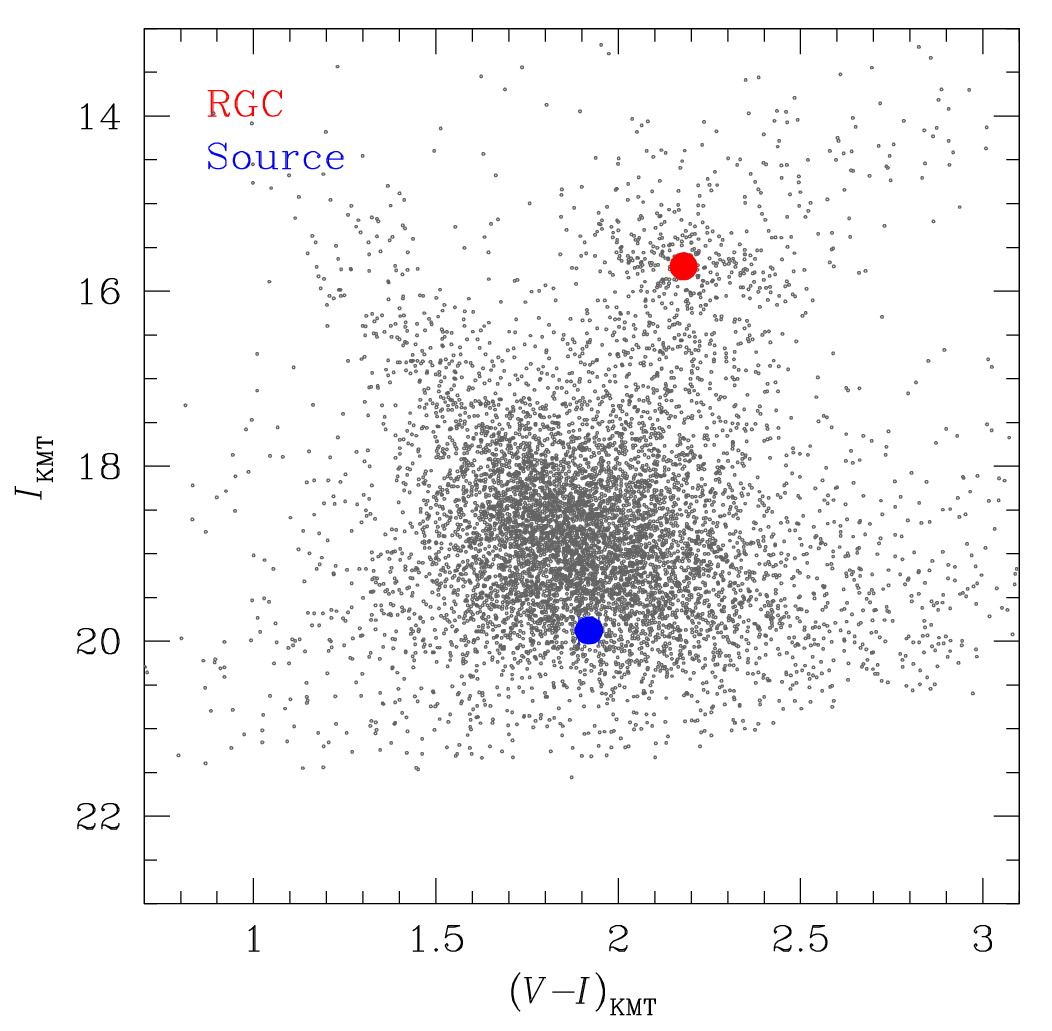}
\caption{
Positions of the source and the centroid of red giant clump (RGC) in the instrumental CMD of 
stars lying near the source.
}
\label{fig:six}
\end{figure}

Figure~\ref{fig:six} shows the positions of both the source and the RGC centroid in the CMD of 
stars located near the source. The measured instrumental color and magnitude of the source are
\begin{equation}
(V-I, I) = (1.919 \pm 0.028, 19.877 \pm 0.012),
\label{eq3}
\end{equation}
and for the RGC centroid,
\begin{equation}
(V-I, I)_{RGC} = (2.179, 15.716).
\label{eq4}
\end{equation}
With the offsets $\Delta(V-I, I) = (-0.260, 4.161)$ and the dereddened values for the RGC centroid,
$(V-I, I)_{{\rm RGC},0} = (1.060, 14.606)$ \citep{Bensby2013, Nataf2013}, the estimated
dereddened color and magnitude of the source are
\begin{equation}
(V-I, I)_0 = (0.800 \pm 0.049, 18.767 \pm 0.023).
\label{eq5}
\end{equation}
These values suggest that the source is a late G-type
main-sequence star located in the Galactic bulge.

Using the measured color and magnitude, the angular radius of the source was determined 
based on the \citet{Kervella2004} relation. 
Because this relation provides a relation 
between ($V-K, V)$ and $\theta_*$, 
we converted the measured $V-I$ color into $V-K$ using 
the color-color relation from \citet{Bessell1988}. The angular source radius estimated 
from this procedure is
\begin{equation}
\theta_* = (0.614 \pm 0.052)~\mu as.
\label{eq6}
\end{equation}
Combining the measured normalized source radius with the angular source radius yields an angular
Einstein radius of
\begin{equation}
\thetae = (0.129 \pm 0.012)~{\rm mas}.
\label{eq7}
\end{equation}
Using this, along with the measured event timescale, the relative lens-source proper motion is
calculated as
\begin{equation}
\mu = {\thetae\over \te} = (4.20 \pm 0.38)~{\rm mas}/{\rm yr}.
\label{eq8}
\end{equation}
The Einstein radii corresponding to the individual lens components $M_1$ and $M_2$ are
$\theta_{{\rm E},1}=[1/(1+q)]^{1/2} \thetae \sim 0.06$~mas and $\theta_{{\rm E},2} = 
[q/(1+q)]^{1/2}\thetae \sim 0.11$~mas, respectively.  These values are notably smaller than 
the typical Einstein radius of around 0.5~mas, which is commonly observed in lensing events 
caused by an M dwarf with a mass of approximately $0.3~M_\odot$ located midway between the 
observer and the source. Moreover, they align with the short event timescales, $t_{{\rm E},1}
\sim 5.3$~days and $t_{{\rm E},2}\sim 9.9$~days.

\begin{figure}[t]
\includegraphics[width=\columnwidth]{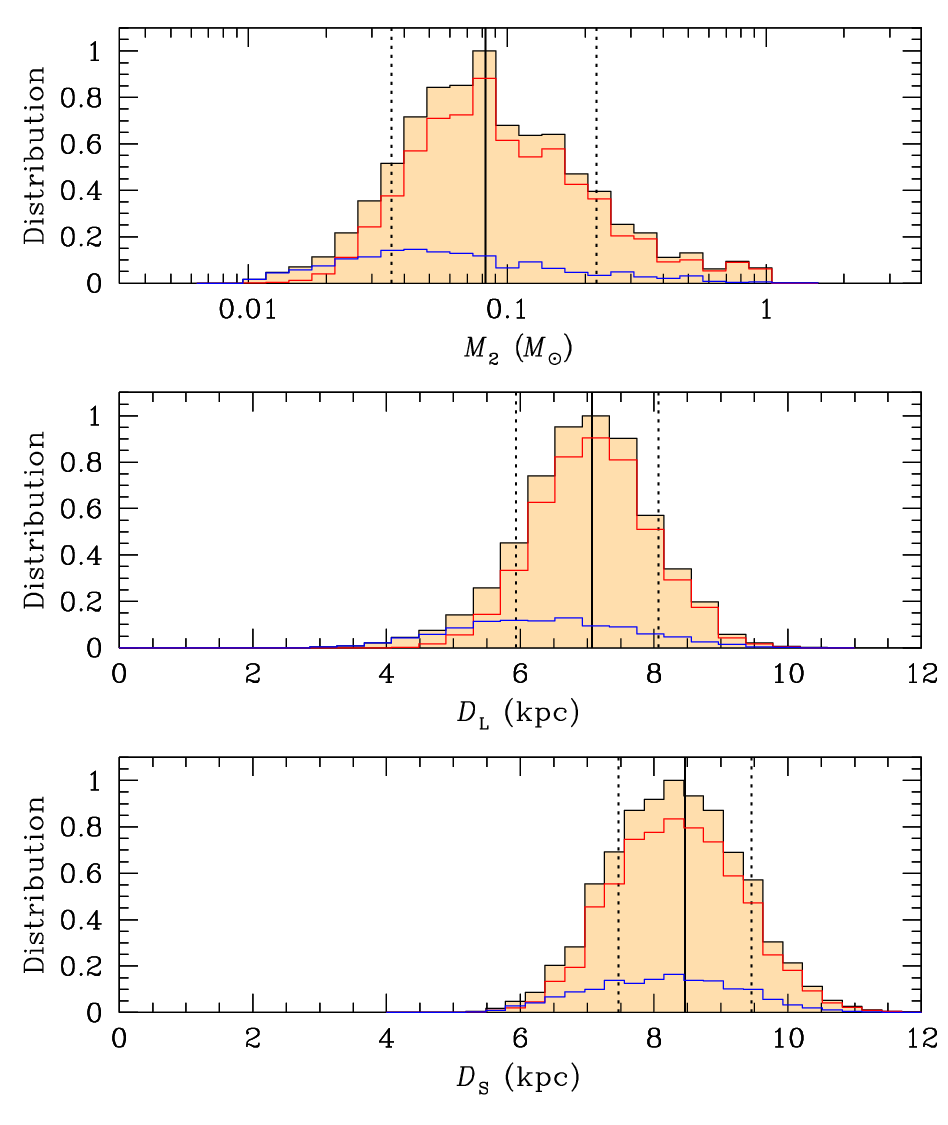}
\caption{
Bayesian posteriors for the mass of the heaviest lens component ($M_2$), distances to the lens 
and source ($\dl$ and $\ds$).  In each distribution, the solid vertical line represents the 
median value, while the 1$\sigma$ uncertainty range is indicated by two dotted lines. The 
curves shown in blue and red represent the contributions from the disk and bulge lens 
populations, respectively, while the black line denotes the combined contribution of both 
populations.
}
\label{fig:seven}
\end{figure}

\section{Physical lens parameters} \label{sec:seven}

We estimate the physical parameters of the lens mass ($M$) and the distance to the planetary 
system ($\dl$) based on the constraints provided by the lensing observables, specifically the 
event timescale and the angular Einstein radius. These observables are related to $M$ and $\dl$ 
through the relations 
\begin{equation}
\te = {\thetae \over \mu}; \qquad
\thetae = \sqrt{\kappa M \pirel }.
\label{eq9}
\end{equation}
Here $\kappa = 4G/(c^2{\rm AU})$, $\pirel={\rm AU}(1/\dl - 1/\ds)$ represents the relative
lens-source parallax, and $\ds$ is the distance to the source.  In addition to these observables, 
the physical lens parameters can be further constrained by the microlens parallax ($\pie$), which 
can be measured from the deviations in the light curve caused by Earth's orbital motion around 
the Sun \citep{Gould1992}. However, for KMT-2024-BLG-0404, this additional lensing observable of 
the microlens parallax could not be measured due to the event's short timescale.

We determined the physical lens parameters through a Bayesian analysis. In the first step of this
analysis, we generated a large sample of artificial lensing events ($6\times 10^6$) using a Monte
Carlo simulation. In this simulation, the lens mass was inferred from a model mass function, and
the distances to the lens and source, along with their relative proper motion, were derived from 
a Galaxy model. For the mass function, we adopted the model from \citet{Jung2018}, and for the
Galaxy model, we used the \citet{Jung2021} model. For each artificial event, we computed the
lensing observables $t_{{\rm E},i}$ and $\theta_{{\rm E},i}$ corresponding to the physical lens 
parameters using the relations in Eq.~(\ref{eq9}). The Bayesian posteriors for the lens mass and 
distance were then constructed by imposing a weight $w_i$ to each artificial event. The weight 
is computed by
\begin{equation}
w_i = \exp\left( -{\chi^2_i \over 2} \right), \qquad
\chi_i^2 = 
{(t_{\rm E} - t_{{\rm E},i} )^2\over \sigma^2(\te)} + 
{(t_{\rm E} - \theta_{{\rm E},i} )^2\over \sigma^2(\thetae)},
\label{eq10}
\end{equation}
where $(\te, \thetae)$ and represent the measured values of the observables and $[(\sigma(\te),
\sigma(\thetae))]$ denote their uncertainties.

The Bayesian posteriors for the lens mass and the distances to both the lens and the source 
are shown in Figure~\ref{fig:seven}.  We also present the posterior for the source distance 
to show the relative positions of the lens and source.  It should be noted that the mass 
posterior corresponds to the most massive lens component, $M_2$. The estimated masses of the 
lens components are:
\begin{equation}
\eqalign{
M_1 = & \ 0.026^{+0.038}_{-0.013}~M_\odot,    \cr
M_2 = & \ 0.090^{+0.133}_{-0.046}~M_\odot,~{\rm and}   \cr
M_3 = & \ 17.3^{+25.5}_{-8.8}~M_{\rm E}.  \cr
}
\label{eq11}
\end{equation}
Here the median was selected as the central value, with the 16th and 84th percentiles of 
the posterior distribution used to define the lower and upper uncertainty bounds, respectively. 
Based on the estimated masses of the lens components, the system is identified as a planetary 
system in which a Neptune-mass planet orbits within a binary consisting of a late M dwarf and 
a brown dwarf. The estimated distance to the lens is
\begin{equation}
\dl = 7.21^{+0.93}_{-0.97}~{\rm kpc}.
\label{eq12}
\end{equation}
The planetary system is most likely located in the bulge, with an 82\% probability.  The 
projected separation between the binary components ($M_1$ and $M_2$) is $a_{\perp, 2} =  
0.85^{+0.11}_{-0.11}$~AU, while the planet is located at a separation of $a_{\perp, 3} = 
0.34^{+0.04}_{-0.05}$~AU from $M_1$.  To ensure dynamical stability in the system, the planet 
should orbit only one of the host stars, and the separation between the binary components must 
be considerably greater than the distance from the planet to its host. This implies that the 
similarity between $a_{\perp, 2}$ and $a_{\perp, 3}$ is likely due to the projection effect.

\section{Summary and conclusion} \label{sec:six}

We have reported the discovery of a new planetary system based on the analysis of the 
lensing event KMT-2024-BLG-0404. The light curve of this event exhibited a complex structure 
with five distinct features, including two prominent caustic spikes. While a binary-lens 
model captured the general anomaly pattern, it could not account for a discontinuous anomaly 
between the two caustic spikes. This remaining anomaly was successfully accounted for by 
introducing an additional lens component with a planetary mass. Based on the estimated masses 
of the lensing components, we identified the system as a planetary system in which a 
Neptune-mass planet orbits a binary composed of a late M dwarf and a brown dwarf. This 
discovery marks the ninth planetary system detected through microlensing where a planet 
orbits a binary host, and it is the first instance where the host consists of a brown 
dwarf and a star.

The detection of planet in a binary system with a brown dwarf component is scientifically 
important because it provides valuable insights into planetary formation and dynamics for 
several reasons.  First, these systems challenge traditional planet formation theories, such 
as core accretion and disk instability, since brown dwarfs may have limited material available 
for planet formation in their disks. As a result, they present unique environments to evaluate 
whether planets can form through standard processes or if alternative mechanisms are required 
in systems with low-mass stars or substellar objects. Second, discovering such planets enhances 
our understanding of the types of objects that can host planets, contributing to the diversity 
of known planetary systems. Finally, the presence of a brown dwarf alongside a planetary system 
in a binary configuration creates a natural laboratory for studying complex gravitational 
dynamics, including orbital resonances, planet-disk interactions, and potential interactions 
between planets and stars.

Planets in binary systems, especially those with brown dwarf companions, are extremely rare. 
The discovery of KMT-2024-BLG-0404L marks the first instance of such a planet identified 
through microlensing. Several analogous systems have been found using radial velocity and 
direct imaging techniques, including HD 41004 \citep{Zucker2003, Zucker2004}, Gliese 229 
\citep{Nakajima1995, Tuomi2014}, and HD 3651 \citep{Fischer2003, Mugrauer2006} through 
radial velocity, as well as 2MASS J0441+2301 \citep{Todorov2010} via direct imaging. Given 
the scarcity of these planetary systems, each new discovery significantly enriches the 
limited sample of known examples. Traditional detection methods like transit or radial 
velocity often struggle to identify brown dwarfs and planets in binary systems due to their 
faintness and low masses. However, microlensing has proven effective in detecting planets 
within these environments, broadening the scope of planetary systems that can be investigated. 
These findings demonstrate that planet formation can occur in previously overlooked settings, 
providing greater insights into the universality of the planet formation process.

\begin{acknowledgements}
This research has made use of the KMTNet system operated by the Korea Astronomy and Space 
Science Institute (KASI) at three host sites of CTIO in Chile, SAAO in South Africa, and 
SSO in Australia.  Data transfer from the host site to KASI was supported by the Korea 
Research Environment Open NETwork (KREONET).  This research was supported by the Korea 
Astronomy and Space Science Institute under the R\&D program (Project No. 2024-1-832-01) 
supervised by the Ministry of Science and ICT.  J.C.Y., I.G.S., and S.J.C. acknowledge 
support from NSF Grant No. AST-2108414.  Y.S.  acknowledges support from NSF Grant No. 
2020740.  W.Z. and H.Y. acknowledge support by the National Natural Science Foundation 
of China (Grant No. 12133005).  W. Zang acknowledges the support from the Harvard-Smithsonian 
Center for Astrophysics through the CfA Fellowship.
\end{acknowledgements}

\begin{appendix}
\section{Scatter plots of MCMC chain points}

\begin{figure}[t]
\includegraphics[width=\columnwidth]{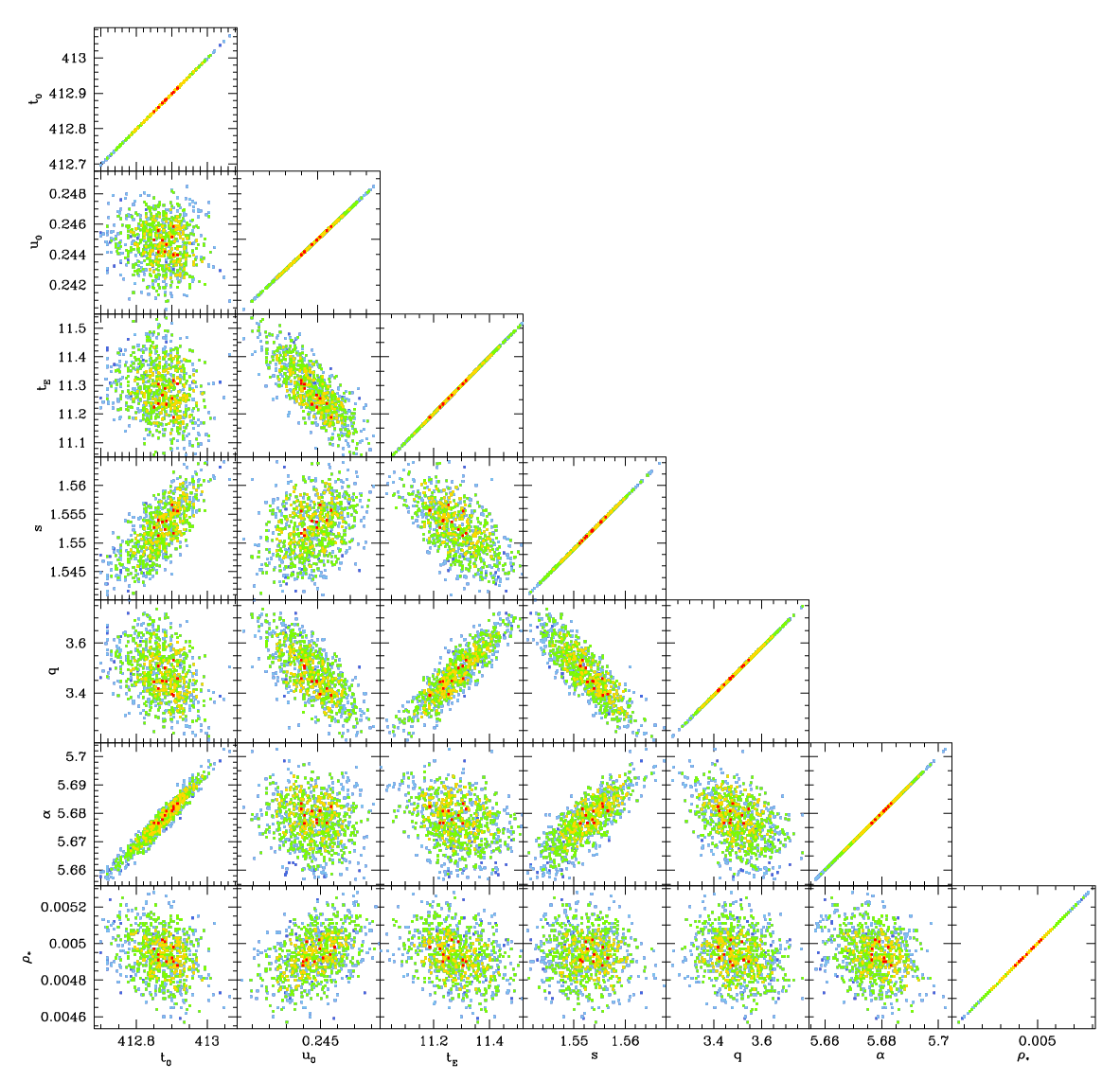}
\caption{
Scatter plot of points in the MCMC chain for the 2L1S model.
Colors are set to represent points within 
$1\sigma$ (red), 
$2\sigma$ (yellow), 
$3\sigma$ (green), 
$4\sigma$ (cyan), and
$5\sigma$ (blue).
}
\label{fig:a1}
\end{figure}

\begin{figure}[t]
\includegraphics[width=\columnwidth]{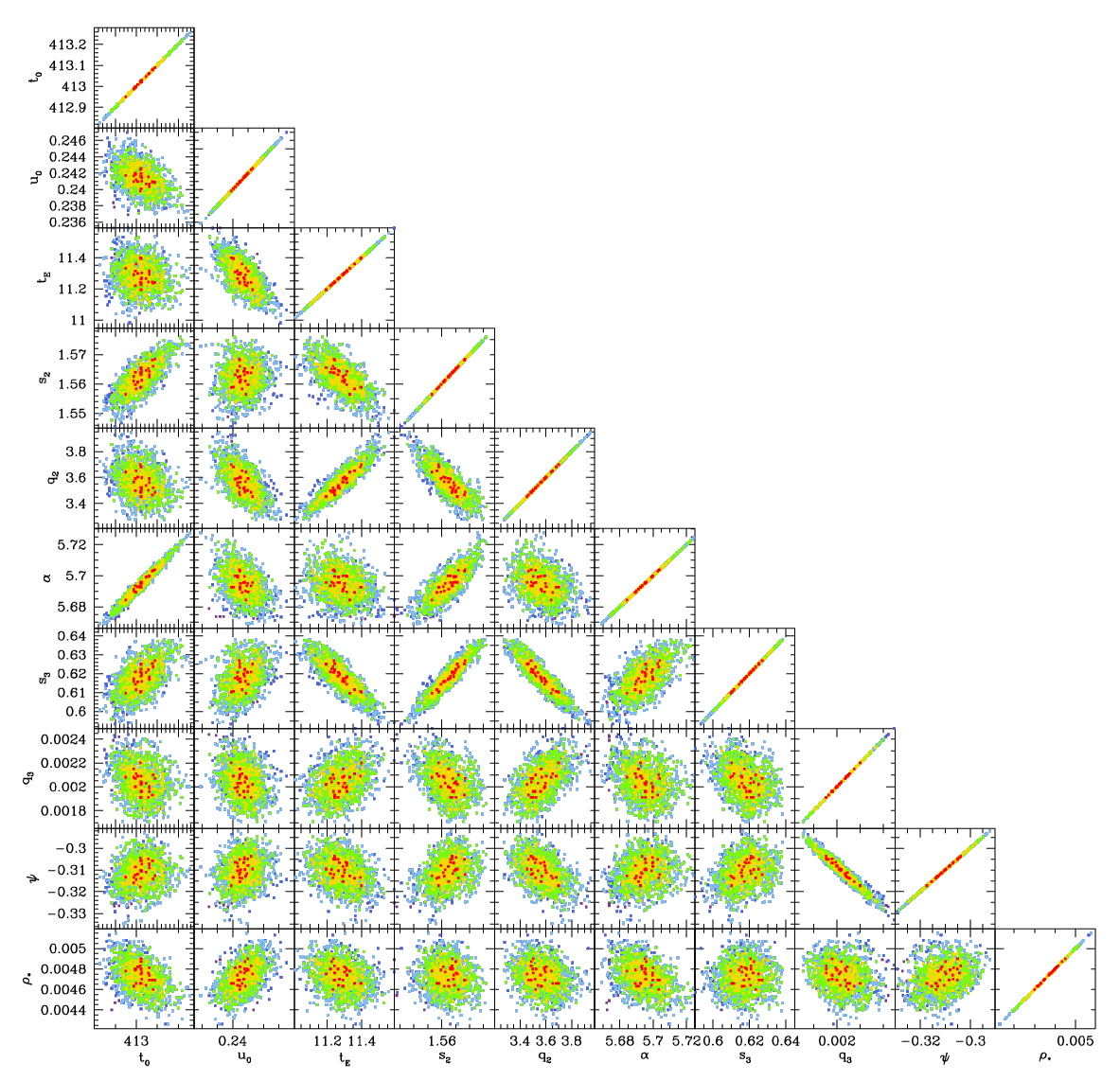}
\caption{
Scatter plot of MCMC chain points for the 3L1S model.
Colors of points are same as those in Fig.~\ref{fig:a1}.
}
\label{fig:a2}
\end{figure}

Figures~\ref{fig:a1} and \ref{fig:a2} display the scatter plots of the Markov Chain Monte 
Carlo (MCMC) chain points corresponding to the 2L1S and 3L1S lensing models, respectively. 
These figures are constructed to illustrate how the lensing parameters are distributed 
throughout the MCMC chains and to highlight the degree of correlation among the parameters 
within each model. In particular, the scatter plots reveal the parameter degeneracies and 
the structure of the uncertainties in the multi-dimensional parameter space, providing 
insight into the nature of the solutions obtained for each model.

\end{appendix}

\begin{thebibliography}{}
\bibitem[Albrow et al.(2000)]{Albrow2000} Albrow, M. D., Beaulieu, J.-P., Caldwell, J. A. R., et al. 2000, \apj, 534, 894
\bibitem[Albrow et al.(2017)]{Albrow2017} Albrow, M. 2017, MichaelDAlbrow/pyDIA: Initial Release on Github,Versionv1.0.0, Zenodo, doi:10.5281/zenodo.268049
\bibitem[Bennett et al.(2016)]{Bennett2016} Bennett, D. P., Rhie, S. H., Udalski, A., et al. 2016, \aj, 152, 125
\bibitem[Bennett et al.(2018)]{Bennett2018} Bennett, D. P., Udalski, A., Han, C., et al. 2018, \aj, 155, 141
\bibitem[Bennett et al.(2020)]{Bennett2020} Bennett, D. P., Udalski, A., Bond, I. A., et al. 2020, \aj, 160, 72
\bibitem[Bensby et al.(2013)]{Bensby2013}  Bensby, T. Yee, J.C., Feltzing, S. et al. 2013, \aap, 549, A147
\bibitem[Bessell \& Brett(1988)]{Bessell1988} Bessell, M. S., \& Brett, J. M. 1988, \pasp, 100, 1134
\bibitem[Dan\u{e}k \& Heyrovsk\'y(2015)]{Danek2015} Dan\u{e}k, K., \& Heyrovsk\'y, D. 2015, \apj, 806, 99
\bibitem[Dan\u{e}k \& Heyrovsk\'y(2019)]{Danek2019} Dan\u{e}k, K., \& Heyrovsk\'y, D. 2019, \apj, 880, 72
\bibitem[Fischer et al.(2003)]{Fischer2003} Fischer, D. A., Butler, R. P., Marcy, G. W., Vogt, S. S., \& Henry, G. W. 2003, \apj, 590, 1081
\bibitem[Gaudi et al.(2008)]{Gaudi2008} Gaudi, B. S., Bennett, D. P., Udalski, A. et al. 2008, Science, 319, 927
\bibitem[Gould(1992)]{Gould1992} Gould, A. 1992, \apj, 392, 442
\bibitem[Gould et al.(2014)]{Gould2014} Gould, A., Udalski, A., Shin, I. -G., et al. 2014, Science, 345, 46
\bibitem[Grieger et al.(1989)]{Grieger1989} Grieger B., K. R., Refsdal S., \& Stabell R. 1989 Abhandlungen aus der Hamburger Sternwarte, 10, 177
\bibitem[Han et al.(2001)]{Han2001} Han, C., Chang, H.-Y., An, J. H. 2001, \mnras, 328, 986
\bibitem[Han(2006)]{Han2006} Han, C. 2006, \apj, 638, 1080
\bibitem[Han(2008)]{Han2008} Han, C. 2008, \apj, 676, L53
\bibitem[Han et al.(2016)]{Han2016} Han, C., Bennett, D. P., Udalski, A., \& Jung, Y. K. 2016, \apj, 825, 8
\bibitem[Han et al.(2017)]{Han2017} Han, C., Udalski, A., Gould, A., et al. 2017, \aj, 154, 223           
\bibitem[Han et al.(2020)]{Han2020} Han, C., Lee, C.-U., Udalski, A., et al. 2020, \aj, 159, 48           
\bibitem[Han et al.(2021a)]{Han2021a} Han, C., Albrow, M. D., Chung, S.-J., et al. 2021a, \aap, 652, A145 
\bibitem[Han et al.(2021b)]{Han2021b} Han, C., Lee, C.-U., Ryu, Y.-H., et al. 2021b, \aap, 649, A91       
\bibitem[Han et al.(2022)]{Han2022} Han, C., Gould, A., Kim, D., et al. 2022, \aap, 663, A145           
\bibitem[Han et al.(2023)]{Han2023} Han, C., Udalski, A., Jung, Y. K., et al. 2023, \aap, 670, A172 
\bibitem[Han et al.(2024)]{Han2024} Han, C., Udalski, A., Jung, Y. K., et al. 2024, \aap, 685, A16       
\bibitem[Jung et al.(2017)]{Jung2017} Jung, Y. K., Udalski, A., Bond, I. A., et al. 2017, \apj, 841, 75
\bibitem[Jung et al.(2021)]{Jung2018} Jung, Y. K., Udalski, A., Gould, A., et al. 2018, \aj, 155, 219
\bibitem[Jung et al.(2021)]{Jung2021} Jung, Y. K., Han, C., Udalski, A., et al. 2021, \aj, 161, 293
\bibitem[Kervella et al.(2004)]{Kervella2004} Kervella, P., Th\'evenin, F., Di Folco, E., \& S\'egransan, D. 2004, \aap, 426, 29
\bibitem[Kim et al.(2016)]{Kim2016} Kim, S.-L., Lee, C.-U., Park, B.-G., et al. 2016, JKAS, 49, 37
\bibitem[Mugrauer et al.(2006)]{Mugrauer2006} Mugrauer, M., Seifahrt, A., Neuh\"user, R., \& Mazeh, T. 2006, \mnras, 373, L3
\bibitem[Nakajima et al.(1995)]{Nakajima1995} Nakajima, T., Oppenheimer, B. R., Kulkarni, S. R., Matthews, K., \& Golimowski, D. A. 1995, Nature, 378, 463-465
\bibitem[Nataf et al.(2013)]{Nataf2013}  Nataf, D. M., Gould, A., Fouqu\'e, P. et al. 2013, \apj, 769, 88
\bibitem[Poleski et al.(2014)]{Poleski2014} Poleski, R., Skowron, J., Udalski, A., et al. 2014, \apj, 795, 42
\bibitem[Rattenbury(2007)]{Rattenbury2007} Rattenbury, N. J., Mao, S., Sumi, T., \& Smith, M. C. 2007, \mnras, 378, 1064
\bibitem[Rhie(2002)]{Rhie2002} Rhie S. H. 2002, arXiv: astro-ph/0202294
\bibitem[Todorov et al.(2010)]{Todorov2010} Todorov, K., Luhman, K. L., \& McLeod, K. K. 2010, \apj, 714, L84
\bibitem[Tuomi et al.(2014)]{Tuomi2014} Tuomi, M., Jones, Hugh R. A., Barnes, J. R., Anglada-Escud\"e, G., \& Jenkins, J. S. 2014, \mnras, 441, 1545
\bibitem[Udalski(2003)]{Udalski2003} Udalski, A. 2003, AcA, 53, 291 
\bibitem[Udalski et al.(2015)]{Udalski2015} Udalski, A., Szyma\'nski, M. K., \& Szyma\'nski, G. 2015, AcA, 65, 1
\bibitem[Udalski et al.(2015)]{Udalski2015b} Udalski, A., Jung, Y. K., Han, C., et al. 2015, \apj, 812, 47
\bibitem[Yang et al.(2024)]{Yang2024} Yang, H., Yee, J. C., Hwang, K.-H., et al. 2024, \mnras, 528, 11
\bibitem[Yee et al.(2012)]{Yee2012} Yee, J. C., Shvartzvald, Y., Gal-Yam, A., et al.\ 2012, \apj, 755, 102
\bibitem[Yoo et al.(2004)]{Yoo2004}  Yoo, J., DePoy, D.L., Gal-Yam, A. et al. 2004, \apj, 603, 139
\bibitem[Zang et al.(2021)]{Zang2021} Zang, W., Han, C., Kondo, I., et al. 2021, RAA, 21, 239
\bibitem[Zhang et al.(2024)]{Zhang2024} Zhang, K., Zang, W., El-Badry, K., et al. 2024, Nature Astronomy, in press 
\bibitem[Zucker et al.(2003)]{Zucker2003} Zucker, S., Mazeh, T., Santos, N. C., Udry, S., \& Mayor, M. 2003, \aap, 404, 775
\bibitem[Zucker et al.(2004)]{Zucker2004} Zucker, S., Mazeh, T., Santos, N. C., Udry, S., \& Mayor, M. 2004, \aap, 426, 695
\end{thebibliography}
\end{document}